\pdfoutput=1
\documentclass{sig-alternate}

\usepackage{graphicx}
\usepackage{xspace}
\usepackage{amssymb}
\usepackage{amsmath}
\usepackage{caption}
\usepackage{subcaption}
\usepackage{paralist}
\usepackage{booktabs}
\usepackage[usenames,svgnames]{xcolor}
\usepackage{pifont}
\usepackage[vlined,ruled]{algorithm2e}
\usepackage{times}
\usepackage{dsfont}
\usepackage{url}

\makeatletter
\g@addto@macro{\UrlBreaks}{\UrlOrds}
\makeatother

\newcommand{\method}{$s$-index\xspace}
\newcommand{\methodr}{$s_r$-index\xspace}

\newcommand{\cmark}{\textcolor{DarkGreen}{\ding{51}}\xspace}
\newcommand{\xmark}{\textcolor{Crimson}{\ding{55}}\xspace}
\newcommand{\dmark}{\textcolor{Peru}{\textbf{--}}\xspace}

\newtheorem{problem}{Problem}

\newtheorem{definition}{Definition}

\begin{document}

\conferenceinfo{WSDM}{'16 San Francisco, CA USA}

\title{s-index: Towards Better Metrics for Quantifying Research Impact}

%
%
%
%
%

\numberofauthors{2}
\author{
\alignauthor
Neil Shah\titlenote{This work was done while working at Microsoft Research Redmond.} \\
       \affaddr{Carnegie Mellon University}\\
			 \affaddr{Pittsburgh, PA}
       \email{neilshah@cs.cmu.edu}
\alignauthor
Yang Song\\
       \affaddr{Microsoft Research Redmond}\\
       \affaddr{Redmond, WA}\\
       \email{yangsong@microsoft.com}
}

\maketitle

\begin{abstract}

The ongoing growth in the volume of scientific literature available today precludes researchers from efficiently discerning the relevant from irrelevant content.  Researchers are constantly interested in impactful papers, authors and venues in their respective fields.  Moreover, they are interested in the so-called recent ``rising stars'' of these contexts which may lead to attractive directions for future work, collaborations or impactful publication venues.  In this work, we address the problem of quantifying \emph{research impact} in each of these contexts, in order to better direct attention of researchers and streamline the processes of comparison, ranking and evaluation of contribution.  Specifically, we begin by outlining intuitive underlying assumptions that impact quantification methods should obey and evaluate when current state-of-the-art methods fail to satisfy these properties.  To this end, we introduce the \method metric which quantifies research impact through influence propagation over a heterogeneous citation network.   \method is tailored from these intuitive assumptions and offers a number of desirable qualities including robustness, natural temporality and straightforward extensibility from the paper impact to broader author and venue impact contexts.  We evaluate its effectiveness on the publicly available Microsoft Academic Search citation graph with over \emph{119 million} papers and \emph{1 billion} citation edges with \emph{103 million} and \emph{21 thousand} associated authors and venues respectively.

\end{abstract}




\section{Introduction}

\label{sec:intro}

The publication and circulation of influential work has served as the cornerstone of research practices since the inception of scientific discovery itself.  Both budding and veteran researchers are known for the quantity and quality of work they produce and moreover by the mark, or \emph{impact}, that they leave on the scientific community.  Furthermore, influential works inspire future endeavors by means of the results and ideas which they put forth -- this phenomenon of incremental discovery is colloquially referred to by the phrase ``standing on the shoulders of giants.''  However, given the continued growth in the sheer amount of literature available today, researchers are deluged with more and more irrelevant information from which they seek only a small fraction.  This makes the measurement of scientific impact of relevant papers, authors and venues an important problem.   

Scientific impact is a central tenet in the evaluation of research success.  While quantitative metrics are not a replacement for carefully reading an author's works and  evaluating peer regard, they are frequently used in practice to provide at-a-glance, summary information. For example, researchers, departments, institutions and venues are commonly evaluated using a number of numerical impact metrics including citation count \cite{yan2011citation}, $h$-index \cite{hirsch2005index} and journal impact factor \cite{garfield1999journal}.  These metrics are typically used for crucial decisions from management perspectives including appointing academic posts, assigning tenure, awarding prizes and electing candidates for prestigious academies.  Similarly, from a researcher's perspective, these metrics can play a major role in determining relevant previous work, future research directions, potential collaborations and strategic publication venues from a citation and recognition perspective.

However, impact quantification is not a trivial problem, particularly because the concept of impact is itself not precisely defined.  Various impact metrics focus on different definitions of impact itself.  For example, paper impact is frequently judged using simple citation count -- thus, papers which have a large quantity of citations are considered to be the most impactful.  Author impact has traditionally been measured using $h$-index, which incorporates (to some extent) both quantity of papers and the quantity of the citations they receive as a measure of quality.  Venue impact is often defined using the impact factor, which considers the average number of citations received per paper over the last 2 years for each venue.  These definitions all inherently capture distinct concepts of impact with inherently different assumptions.  Hence, it is unsurprising that current state-of-the-art metrics are not without a number of inconsistencies and unexpected behaviors in practice.  In fact, numerous works from different fields including social sciences, bibliometrics and physics establish notable problems with evaluation arising from the use of these metrics in practice \cite{wilhite2012coercive, bartneck2010detecting, plos2006impact, cheng2006manipulability}.  We argue that given the career and livelihood ramifications of these metrics, it is important for impact metrics to be principled and in-line with human understanding.  In this work, we focus exactly on this problem of developing improved, powerful and practical metrics for effectively quantifying research impact.  

\begin{figure}[t!]
\centering
\includegraphics[width=0.7\linewidth]{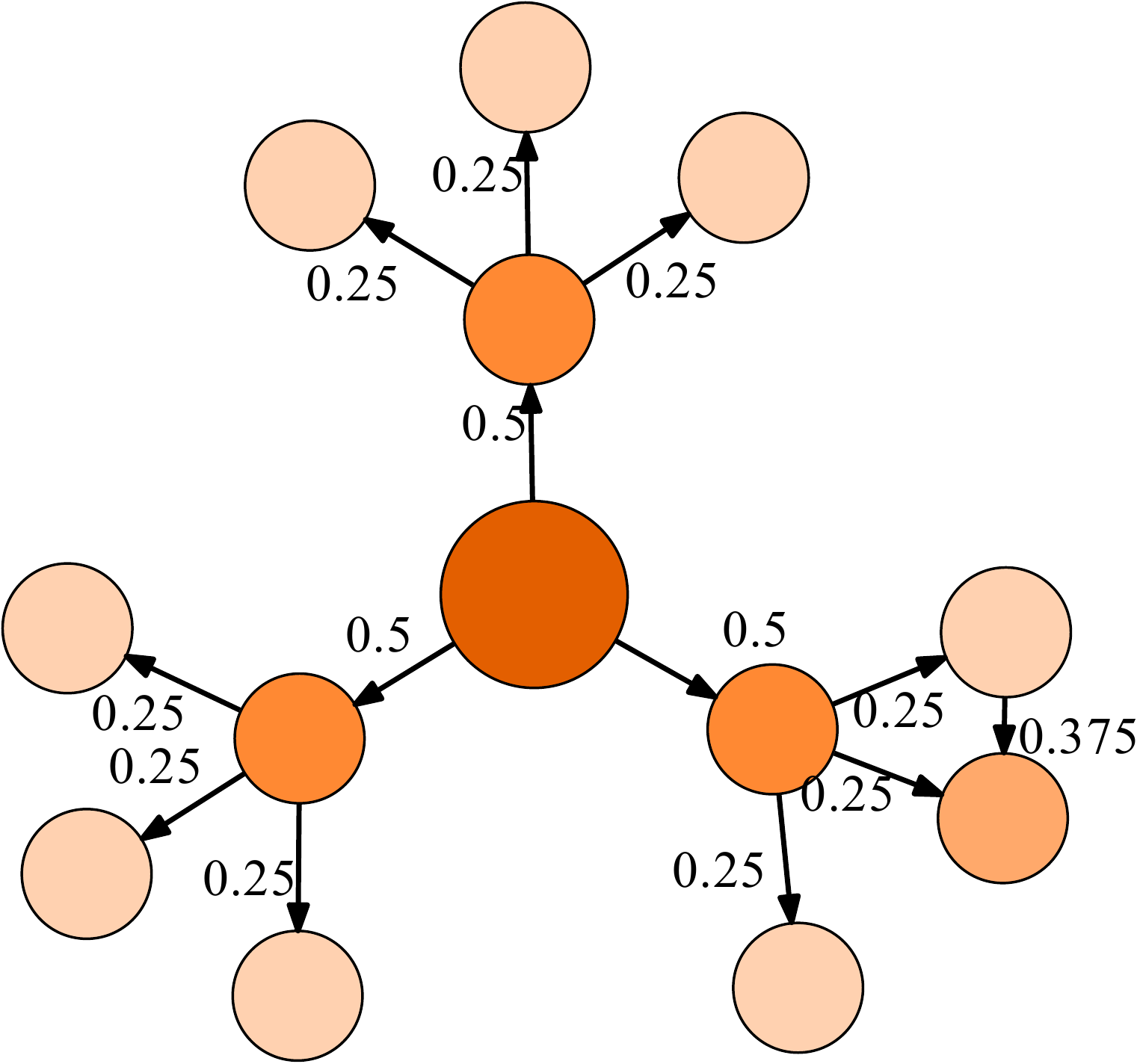}
\caption{\method measures entity impact by modeling influence propagated by papers over citation networks.  Edge $p_1 \rightarrow p_2$ denotes that $p_2$ cites $p_1$ and edge weights denote influence from the center (source) node reaching a receiver.}
\label{fig:concept} 
\end{figure}

We begin by identifying a number of desiderata that good impact metrics should obey.  These attributes have firm grounding in human intuition about how impact should be manifested by different entities (in this work, we consider impact of papers, authors and venues).  We next identify relevant prior works and current state-of-the-art metrics used to quantify impact in practice and evaluate when these metrics fail to satisfy intuitive properties.  In response, we build necessary groundwork for and propose the \method metric which is designed to exhibit these traits.  Our approach computes paper impact by modeling influence propagated by papers over a paper-paper citation network (see Figure \ref{fig:concept} and extends this principle to associated author and venue nodes in a heterogeneous paper-author-venue citation network.  \method offers numerous comparative advantages over existing state-of-the-art impact metrics and emphasizes both \emph{quantity} and \emph{quality} of research while adhering to a number of important properties related to the tradeoff between the two.  

Our work offers a number of notable contributions towards solving the problem of research impact quantification:
\vspace{1mm}
\begin{compactenum}
\item \textbf{Analysis:} We identify a number of features that impact metrics should obey in order to be employed in practice, and analyze when current state-of-the-art metrics do not perform accordingly.
\item \textbf{Algorithm:} We build intuition for and propose a fast and scalable algorithm to compute the \method metric, which quantifies research impact based on influence propagation over a citation graph and is currently deployed at Microsoft.
\item \textbf{Evaluation:} We evaluate the \method on the large, Microsoft Academic Search citation graph with over \emph{119 million} papers, \emph{1 billion} citation edges, \emph{103 million} authors and \emph{21 thousand} venues and show promising results in practice.  
\end{compactenum}
\vspace{1mm}

\textbf{Reproducibility:} The Microsoft Academic Search data used in our work is freely available at \url{research.microsoft.com/en-us/projects/mag/}.  Both MATLAB and MS-SQL relational implementations of our algorithm are made available at \url{cs.cmu.edu/~neilshah/code/sindex.tar.gz}.

\section{Proposed Desiderata}

\label{sec:desiderata}

In order to build an improved impact metric, it is important to identify how we expect impact to be defined in practice.  The Economic and Social Research Council (ESRC) defines academic research impact as follows:

\begin{quote}

\textbf{Academic impact} is the demonstrable contribution that excellent social and economic research makes to scientific advances, across and within disciplines, including significant advances in understanding, method, theory and application. \cite{impactdefn} 

\end{quote}

Beyond this broad definition, there are few real constraints on how research impact is construed.  In practice, impact is traditionally defined in some means by using available information about \emph{citations}, or references which publications (henceforth referred to as papers) make between each other to signify drawing from, or leveraging other works in their own.  In their simplest form, citations are direct links which convey the number of ``uses'' of a paper.  However, simply tallying the number of citations received by various papers, authors and venues makes for a very elementary impact metric, which can only be considered a first-order approximation of the notion of ``contribution'' referred to in the above definition.  In reality, impact is both a direct and indirect phenomenon, and metrics which quantify impact should account for this property, amongst several others.  

In this section, we pose the question: what makes a good impact metric?  To answer it, we propose and define these desiderata which should be considered in designing and employing the use of an impact metric.  They are (a) \emph{volume sensitivity}, (b) \emph{prestige sensitivity}, (c) \emph{robustness}, (d) \emph{extensibility}, (e) \emph{temporality}, (f) \emph{interpretability} and (g) \emph{computability}.  While we discuss the properties in the context of paper impact, the principles extend naturally to the broader author and venue contexts.

\subsection{Volume Sensitivity}

\emph{Volume sensitivity} reflects the concept that the more a work is cited, the more impactful it is.  This does not imply that direct citation count is the best impact metric, but rather that between two otherwise equivalent papers $a$ and $b$, $a$ is more impactful if it has an extra citation over $b$.  Thus, more citations does not negatively affect impact and only helps it.  This property is intuitive and is the fundamental tenet of citation counting for impact.

\subsection{Prestige Sensitivity}

\emph{Prestige sensitivity} captures the idea that impact metrics should weigh citations from different papers differently.  In other words, not all citations are considered equally.  The intuition behind this property is that citations from ``high-quality'' papers should matter more than those from ``low-quality'' ones (for example, a widely-acclaimed seminal work in a famous journal versus an uncited and unpublished work posted online).  Typically, the quality of citing papers is defined recursively in some fashion by their own citation count.  Most traditionally used impact metrics offer no or very limited prestige sensitivity.  

\subsection{Extensibility}

\emph{Extensibility} refers to the idea that the principles by which an impact metric is defined should extend to the impact definitions for other entities.  Specifically, the property of extensibility ensures that impact is defined in a unified and coherent way across papers, authors and venues rather than a segmented one which only applies to certain entities.  It is both perplexing and unintuitive that currently used state-of-the-art impact metrics convey effectively \emph{different} measures of impact for papers, authors and venues.  We argue that papers are the fundamental building block of impact, and authors and venues are simple aggregates of associated papers.  An author cannot have impact without the papers which he writes, nor can a venue have impact without the papers which it publishes to the scientific community at large.  

\subsection{Temporality}

Impact can be viewed in two ways.  Firstly, impact can be considered as a static, ``lifetime achievement award'' which quantifies total influence from inception onwards.  This could be considered as the total impact of a paper from publication or the total impact of an author or venue from its first publication.  Secondly, impact can be examined as a dynamic, ever-changing property which quantifies influence \emph{recently}.  For example, a paper written in the 1600s may have a large static impact, but the dynamic impact may wane over time due to a variety of reasons, including declining interest in the specific work, the field or as a result of shift towards newer results.  We argue that while impact metrics should be extensible across entity contexts, they should also be extensible over time and offer dynamic counterparts with similar principles -- we refer to this idea of extensibility over time as \emph{temporality}.

Impact metrics which offer temporality are particularly useful because they can measure (by definition) how influence changes over time, and thus offer means of measuring recent productivity and popularity.  This is especially useful for comparison purposes between papers, authors and venues at different points in their respective careers.  Furthermore, temporality is associated with predictability, as those with higher recent impact are intuitively expected to coincide with so-called ``rising stars.''

\subsection{Interpretability}

Metrics can be arbitrarily simple or complex.  Often, more complex models which rely on multiple data sources offer more expressibility than simpler ones.  In the impact metric context, one can consider the simplest metric to be citation count.  However, a more complicated (though perhaps rich) metric might account for a variety of factors such as auxiliary author/institution/venue features, semantic similarity with previous works, or produce tuples instead of single numbers.  However, one must keep in mind that impact metrics are meant to be used by \emph{humans} -- thus, they must offer \emph{interpretability}.  In practice, these metrics are used not only to rank, but also to compare and predict.  It is of utmost importance that those who use them have some understanding of how the concept of impact is being ranked, and what they are comparing and predicting.    

\subsection{Robustness}

When a metric is introduced as a quantifier, its value as a measure immediately begins to decline.  This consequence was initially formulated in the economic context and is known as Goodhart's Law, which states ``when a measure becomes a target, it ceases to be a good measure,'' in reference to the response of investors to act in ways which they seek to benefit from economic regulations.  Analogously, as impact metrics are proposed, researchers will seek to adopt practices which enable them to capitalize from the metric and better their rankings.  Thus, it is important that impact metrics are \emph{robust}, or difficult to rig or game by means of disreputable practices (such as unwarranted self-citation, double publication and citation trading).  

In practice, self-citation can be used both legitimately and illegitimately and it can be difficult to automatically discern between the two.  Thus, we argue that impact metrics which explicitly penalize self-citation are ideal.  Rather, it is more promising to measure impact in a way which diminishes the incentive to self-cite illegitimately -- note that this is inherently impossible with metrics that prestige sensitivity and treat all citations equally.

\subsection{Computability}

Good impact metrics should be easily \emph{computable}.  Citation networks and their more complex, heterogeneous representations are constantly growing with the volume of available literature.  Impact metrics which are impractically expensive or difficult to compute, no matter how expressive or even interpretable they are, are simply not practically usable.  Online citation and ranking databases which are commonly used such as Microsoft Academic Search \cite{mas}, Google Scholar \cite{gscholar}, CiteSeerX \cite{citeseerx} and ArnetMiner \cite{aminer} deal with very large datasets and require frequent updates to impact metrics given the frequency with which they index new articles -- thus, requiring several days or longer to compute metric scores is an unattractive option.  Computability is an especially important consideration for complex metrics which incorporate costly operations such as semantic similarity and content-based approaches, centrality metrics and slow-converging graph algorithms.  Furthermore, content-based approaches which use topic modeling or other randomized algorithms are approximate, meaning that impact metrics can be computed substantially differently even on the same dataset -- this is, of course, undesirable. 

\section{Prior Work \& Analysis}

{\setlength{\tabcolsep}{6pt}
\begin{table*}[ht!]
\small
\centering
\caption{Qualitative comparison with modern research impact metrics.}
\label{tbl:compprev}
\begin{tabular}{lccccccc}
\toprule
               & \textbf{Volume Sensitive} & \textbf{Prestige Sensitive} & \textbf{Extensible} & \textbf{Temporal} & \textbf{Interpretable} & \textbf{Robust} & \textbf{Computable} \\
\midrule \midrule
Citation count &        \cmark            &   \xmark   &       \cmark      &   \cmark      &   \cmark   &      \xmark     &    \cmark      \\
$h$-index      &        \dmark            &      \dmark      &  \xmark     &    \cmark     &    \cmark      &  \xmark       &  \cmark        \\
JIF            &        \cmark            &   \xmark   &       \cmark      &   \cmark      &   \cmark   &      \xmark     &    \cmark      \\
PageRank       &        \cmark            &   \dmark        &  \cmark       &  \cmark      &   \dmark       &   \cmark     &   \dmark            \\
\midrule
\method        &         \cmark           &  \cmark          &  \cmark       &  \cmark       & \cmark        &   \cmark     &  \cmark             \\
\bottomrule
\end{tabular}
\end{table*}

\label{sec:pwork}

\subsection{Prior Work}

\subsubsection{Impact Metrics}

Citation count is perhaps the oldest and most commonly used metric for measuring research paper impact.  \cite{beel2009google} notes that Google Scholar considers citation count to be the highest weighted factor for paper ranking.  Ranking by citation count is a common, but controversial practice in that it reinforces the rich-get-richer concept (Matthew effect).  \cite{walker2007ranking} proposes the CiteRank algorithm for paper ranking, which is similar to Google's PageRank algorithm \cite{ipsen2006mathematical} but distributes random surfers exponentially with age, favoring more recent works.  This assumption biases against older papers, which is an unintuitive assumption for overall impact calculation.  \cite{chen2007finding} also uses PageRank to assess importance of papers published in Physical Review journals.

\cite{hirsch2005index} proposes the $h$-index to quantify author's research output.  To compare researchers with different career lengths, the $m$ quotient, derived from dividing$h$ by the length of the author's academic career, is also suggested.  \cite{egghe2006theory} proposes the $g$-index as an alternative to more heavily account for an author's top contributions, which may have disproportionately more citations than his less popular papers.  The $a$, $r$ and $ar$-index defined in \cite{jin2006h} and \cite{jin2007r} use variants of mean citations of popular papers in the \emph{Hirsch core} to capture the average impact of an author's high-performing papers in order to less penalize authors with high $h$-index.  Google Scholar recently introduced the $i10$-index \cite{i10gscholar}, defined as the number of papers with 10 or more citations.  \cite{bornmann2008there} uses factor analysis to classify these indices into two main types which emphasis work quantity and quality.  These groups represent the concepts of \emph{volume sensitivity} and \emph{prestige sensitivity}, respectively.

Venue impact of journals in the same field is usually computed using journal impact factor (JIF) \cite{garfield2006history}.  However, JIF computes a mean over a heavy-tail distribution of citation counts and is thus of limited value as a statistical measure.  \cite{bergstrom2008eigenfactor} describes the EigenFactor metric for ranking journals based on PageRank on journal-journal citation graphs generated through network inference via paper citation.  

\subsubsection{Impact Prediction}

A number of works utilizing regression and classification have been proposed in the past with the aim of predicting citation count or otherwise quantifying research success.  Though our work does not directly focus on prediction, these works relate to the concept of \emph{temporality} for impact metrics and are thus described.  \cite{manjunatha2003citation} proposes $k$ nearest neighbors (KNN) regression on citation count differences over previous years for the KDD Cup 2003 citation prediction task.  \cite{shi2010citing} identifies features distinguishing well and poorly cited papers on local \emph{reference networks} of paper-paper citation graphs.  \cite{livne2013predicting, yan2012better} use various regression techniques including support-vector regression (SVR) and linear regression (LR) on numerous features to predict field-specific paper citation count several years ahead.  \cite{dong2015will} uses multiple classification models to determine whether a given paper will increase the author's $h$-index or not.  \cite{wang2013quantifying} identifies some important mechanisms that play a role in long-term paper citation count including aging and the Matthew effect.

\subsection{Analysis}

In this section, we qualitatively evaluate the performance of current state-of-the-art impact metrics which are commonly used today.  Given the breadth of the previously described prior work, we select 4 representative approaches which sufficiently span the multitude of approaches.  They are (a) citation count, (b) $h$-index, (c) JIF and (d) PageRank.  Table \ref{tbl:compprev} gives a high-level summary of the strengths and weaknesses of each approach with respect to the desiderata identified in Section \ref{sec:desiderata}.  For computability results, we assume a heterogeneous citation graph $G$, defined as follows:

\begin{definition}[Citation graph $G$]
$G$ has $|P|$ paper nodes, $|A|$ author nodes, and $|V|$ venue nodes, $|E_{pp}|$ paper-paper edges, $|E_{pa}|$ paper-author edges and $|E_{pv}|$ paper-venue edges where edge $p_1 \rightarrow p_2$ denotes that paper $p_1$ is \underline{cited by} paper $p_2$, $p \rightarrow a$ denotes that paper $p$ is \underline{authored by} author $a$ and $p \rightarrow v$ denotes that paper $p$ is \underline{published by} venue $v$.
\end{definition}

\subsubsection{Citation count}

Citation counting involves tallying the number of citations received by a paper, author or venue.  We will denote the number of citations of a paper $p$ by $C(p)$.  

It is a purely \emph{volume sensitive} metric, and offers no means of \emph{prestige sensitivity}, since all citations are weighted equally in computing impact.  In many cases, citation count does not correspond to the supposed impact of a paper -- for example, \cite{aminer} shows substantial differences between ``best papers'' from computer science conferences versus the most cited ones.  This can happen for numerous reasons: in the case of ubiquitously used results in which the original work is no longer cited, in the case of incremental works leading up to an important result or in the case of older papers which are buried by new literature.  Citation counting is \emph{extensible}, as it can be applied to authors and papers quite easily.  Furthermore, it is \emph{temporal} given appropriate conditioning on input data.  It offers straightforward \emph{interpretability} as the number of citations of an entity.  However, it is not \emph{robust} given that it is highly susceptible to disreputable practices such as self-citation, double publication and citation trading -- \cite{aksnes2003macro} shows that self-citation makes up a significant part of general citation activity and its presence plays a substantial role in citation counts of papers and authors.  Citation counting is easily computable and can be computed in $O(|E_{pp}|)$ for all papers in the paper-paper citation graph $G$.

\subsubsection{$h$-index}

The $h$-index of an author $a$ with published papers $P^a$ is defined as 
\begin{equation*}
H(a) = \max_{h \in \mathbb{N}^+} h \;\; \mbox{s.t.} \;\; \left( \sum\limits_{p \in P^a} \left[ C(p) \geq h \right] \right) \geq h
\end{equation*}
\noindent where $\left[ \, \cdot \, \right]$ serves as an indicator function.  Informally, it is referred to as the maximal $h$ for which the author has $h$ papers with $h$ or more citations each.  

$h$-index is only somewhat \emph{volume sensitive} and \emph{prestige sensitive} despite considering some concept of both paper quantity and quality.  To illustrate, let us consider two scientists $a$ and $b$ with varying publication records.  Suppose that both $a$ and $b$ have published 10 papers with 10 citations each, but $b$ has additionally published 90 papers which received 9 citations each.  Counterintuitively, both scientists have an equivalent $h$-index of 10, despite scientist $b$'s much higher quantity of work.  Alternatively, consider scientists $c$ and $d$ who have both published 5 papers.  However, each of $c$'s papers has 5 citations each, whereas each of $d$'s papers has 500 citations each.  Once again, both authors have an equivalent $h$-index of 5, despite scientist $d$'s much higher quality of work.  $h$-index is defined only in the author context (though it is sometimes used for venue impact), but offers no extensions for paper impact and is thus not \emph{extensible}.  It is however \emph{temporal} given appropriate conditioning on input data.  It has an \emph{interpretable} definition as well.  \cite{schreiber2007self} and \cite{bartneck2010detecting} show that strategic self-citation can dramatically boost $h$-index over time due to the metric's equivalent treatment of self-citations and citations from others, thus limiting \emph{robustness}.  $h$-index is relatively \emph{computable} and can be computed in roughly $O(|E_{pp}| \, + \, |E_{pa}| \, + \, |A| \, w \,log(w))$ for $|A|$ authors and $w$ mean paper-author edges (papers per author).

\subsubsection{Journal Impact Factor}

The JIF of a venue $v$ with papers published in the previous 2 years $P_2^v$ is defined as 
\begin{equation*}
J(v) = \frac{\sum\limits_{p \in P_2^v} C(p)}{|P_2^v|}
\end{equation*}
\noindent Thus, it is computed as the mean number of citations received by papers published in that time frame.  

JIF shares similar properties to citation counting given its innate dependent on the citation count.  It is \emph{volume sensitive} but not \emph{prestige sensitive}.  It is \emph{extensible} as citations can be agglomerated on a paper, author or venue context (though should not be used in practice outside of the venue context \cite{easeinappropriate}) .  It is additionally \emph{temporal} given mean computation over the last $t$ years (though $t=2$ in practice, $t \in \mathbb{N}$ could be used more generally).  JIF is also \emph{interpretable}.  However, it is not \emph{robust} both due to sole emphasis on citation quantity as well as due to coercive self-citation and impact factor boosting tricks mentioned in \cite{wilhite2012coercive} and \cite{plos2006impact}.  Furthermore, given that JIF computes a mean over a power-law distribution, it is highly susceptible to ``black-swan'' outliers -- for example, the impact factor of the journal \emph{Acta Crystallographica} rose from 2.05 to 49.93 in 2009, more than \emph{Nature} and \emph{Science} due to the result of just 1 publication \cite{reutersoutlier}.  JIF is easily computable and can be computed in $O(|E_{pp}| \, + \, |E_{pv}|)$ for all journals.

\subsubsection{PageRank}

\label{pwork:pr}

The PageRank $PR(p)$ of paper $p$ in $G$ is defined as
\begin{equation*}
PR(p) = \frac{1-d}{|P|} \, + \, d \sum\limits_{q \, \in \, L^{-1}(\{p\})} \frac{PR(q)}{L(\{q\})}
\end{equation*}
\noindent where $d$ is a damping factor in $(0,1)$, $L^{-1}(\{p\})$ denotes $p$'s references and $L(\{p\})$ denotes those papers which $q$ is cited by.  It can also be computed as the dominant eigenvector of the associated stochastic Google matrix described in \cite{ipsen2006mathematical}. 
	
PageRank is \emph{volume sensitive} since the more citations a paper $p$ has, the higher its PageRank in otherwise equivalent contexts.  However, PageRank has limited \emph{prestige sensitivity} -- although papers with many citations pass greater influence to their own references, PageRank has the property that a paper's propagated influence is apportioned \emph{equally} between its references.  This means that a citation from a paper with many references is less important than one from a paper with fewer references.  In the web-context in which PageRank was originally proposed, this assumption makes sense given that pages which have many external links are often link farms or low-quality web-indices.  However, in the citation graph context, the length of a work's reference list is not a measure of \emph{exclusivity} but rather of \emph{relevance} -- it is certainly not obvious that the length of the reference list of a paper bears any influence on the quality of the work itself.  In fact, some of the works with the highest reference counts are textbooks, surveys and tutorials which are highly impactful in making technical expertise accessible to many authors.  Thus, we argue that this exclusivity property of PageRank is ill-suited to represent prestige in citation graphs.  PageRank is \emph{extensible} given appropriate network inference to construct author-author and journal-journal citation graphs.  It is also \emph{temporal} given appropriate conditioning on input data.  Though PageRank in web contexts has the traditional ``random surfer'' interpretation, the model lacks \emph{interpretability} as a metric in the research impact context given the exclusivity assumption.  Furthermore, given numerous issues of \emph{computability}, including the many iterations required for sufficient convergence in practice despite $O(|E_{pp}|)$ runtime per iteration and generally inadequate machine precision due to $|P|$, the resulting $(0,1]$ scores from PageRank computation are difficult to interpret in the research impact context.  

\section{Proposed s-index}
\label{sec:proposedmethod}

In this section, we first build intuition towards, and next define the \method for quantifying paper, author and venue impact.  Lastly, we give a scalable algorithm for computing \method efficiently.

\subsection{Intuition}
\label{sec:intuit}

We begin by posing the following problem:

\begin{problem}[Paper impact] 
{\bf Given:} the citation graph $G$, {\bf find:} a metric by which to quantify the value of papers according to their research impact.
\end{problem}

We start from the most fundamental idea: counting the number of citations of each paper in order to rank them is a straightforward first-order approximation of any \emph{volume sensitive} impact metric.  Citation count in some sense can be construed as the immediate usefulness of the paper to other researchers.  If paper $p_1$ is cited by paper $p_2$, we take this to mean that $p_2$ has derived some useful information from $p_1$ -- in other words, $p_1$ \emph{influenced} $p_2$.  However, we know that citation count focuses only the quantity of citations, but places no emphasis on the quality.  

To incorporate \emph{prestige sensitivity}, we can then examine the citations which paper $b$ received.  For ease of explanation, we will now define a function $L(H)$ which, given a subset of papers $H \subseteq P$, will return the maximal subset of papers $T \subseteq P$ which cite some paper $h \in H$ -- in other words, there exists an edge $e \in E_{pp}$ from $h \rightarrow t$ for some $h \in H$ and $t \in T$.  These are referred to as the \emph{descendants} of $H$.  We additionally define $L^k(H)$ for $k \in \mathbb{N}$ which denotes $k$ compositions of $L$ -- that is, $L^1(H) = L(H)$, $L^2(H) = L(L(H))$ and so on.  It will also be useful to define behavior for $k \leq 0$.  $k = 0$ indicates 0-step neighbors -- this means that $L^0(H) = H$.  For $k < 0$, we consider the \emph{ancestors} of the nodes in $H$ rather than the \emph{descendants}, such that $L^{-1}(H)$ returns the maximal subset of papers $T \subseteq P$ which are cited by some paper $h \in H$ -- in other words, there exists an edge $e \in E_{pp}$ from $t \rightarrow h$ for some $t \in T$ and $h \in H$.  The compositional behavior of $L^k(H)$ for negative $k$ is defined similarly: $L^{-2}(H) = L^{-1}(L^{-1}(H))$ and so on.  Note that we refer to proximal links as ancestors and descendants rather than in-links and out-links to clarify that the paper-paper graph is not just directed, but also effectively acyclic -- reciprocal citation relationships are extremely rare given the temporal connotation of citation, and only possible with citation to a paper published in the future.

In our example so far, $L(\{p_1\}) = \{p_2\}$, and presuming that $p_2$ only cites $p_1$, $L^{-1}(\{p_2\}) = p_1$.  If $p_2$ has received a large number of citations, then we can consider $p_2$'s citation to $p_1$ as more valuable than, for example, a citation from paper $p_3$ which also cites $p_1$ but itself has fewer citations.  Intuitively, we use $p_2$'s quantity of citations as a measure of the quality of $p_2$'s citation to $p_1$.  Thus, we examine $p_1$'s 2-step descendants $L^2(\{p_1\})$, instead of the 1-step descendants $L(\{p_1\})$ as for simple citation count.  In fact, we can take yet another step and look at the 3-step descendants $L^3(\{p_1\})$ in order to gain further confidence in the quality of papers in $p_1$'s 1-step and 2-step descendants, and so on. 

Moreover, just as $p_1$ influenced $p_2$ (more generally $L(\{p_1\})$), we can consider that $p_1$ \emph{also} influences the papers which cite $p_2$ (more generally $L^2(\{p_1\})$).  This is because those papers which cite $p_2$ have indirectly drawn some useful information from $p_1$ by means of $p_2$.  One can imagine that the more steps away from $p_1$ a paper is in $G$, the less it draws from, or is influenced by $p_1$.  Thus, the walk length and influence should be inversely correlated.  Given the nature of the problem, we conjecture that a constant fraction of the influence will wane per each further step away from $p_1$ -- thus, the influence should decay exponentially with respect to path length.  This assumption is in line with the damping factor idea used in PageRank.  Though we could consider arbitrary walk lengths from each node, spanning descendants as far as the full diameter of $G$, it is more intuitive to consider shorter lengths in practice given that the proportion of influence decays rapidly from $a$ to $b$ as the walk between the two becomes longer.  Note that we \emph{do not} weight influence by the inverse of the number of 1-step ancestors of $p_2$ $L^{-1}(\{p_2\})$, as this idea is characteristic of PageRank.  We discuss in Section \ref{pwork:pr} the reasons for which this notion is ill-suited for the impact context.

Given that there are potentially arbitrarily many walks from $p_1$ to $p_2$, which should we choose?  One option is to choose the shortest -- this way, the shorter the shortest walk, the more $p_1$ influences $p_2$.  Although this idea is reasonable, we can in fact leverage an even better option which is even more expressive.  Figure~\ref{fig:feedfwd} shows the case of a ``feed-forward loop,'' in which paper $p_1$ is cited by both $p_2$ and $p_3$, but $p_2$ is also cited by $p_3$, which makes the weakness of the shortest walk idea more apparent.  Specifically, this notion conveys that $p_1$ influences $p_2$ and $p_3$ equally, given that the shortest walks to both papers are of length 1.  However, given that in addition to citing $p_1$ directly, $p_3$ also cites $p_2$ which cites $p_1$, we can say that $p_1$ influences $p_3$ both directly and by proxy via $p_2$, but only influences $p_2$ directly.  In practice, this interaction further substantiates the influence that $p_1$ has on $p_3$, suggesting that $p_3$ has been more influenced by $p_1$ than $p_2$.  Thus, making use of multiple walks between papers and the interactions they represent is a promising approach for defining a more powerful measure of influence.   

\begin{figure}[t!]
\centering
\includegraphics[width=0.7\linewidth]{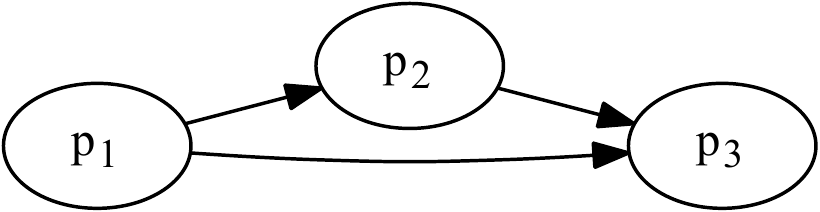}
\caption{A ``feed-forward loop'' in which paper $p_1$ is cited by both $p_2$ and $p_3$, but $p_2$ is also cited by $p_3$.}
\label{fig:feedfwd} 
\end{figure}

This concept of influence propagating between papers is characteristic of the phrase ``standing on the shoulders of giants.''  With this intuition, we express that if paper $a$ influences paper $b$, $a$ should get some credit for $b$'s successes.  This concept enforces \emph{robustness} as self-citation and citation trading practices become far less impactful in comparison to producing highly influential work which can enjoy exponential \method growth.  This is exactly the notion that \method is built on.  

\subsection{Definition}
\label{sec:defn}

With the established concepts from Section \ref{sec:intuit}, we define the \method $\mathcal{S}$ of a paper $p$ as
\begin{equation*}
\mathcal{S}(p) = \sum_{i=1}^{m} \, I(i) \, W(p, \, L^i(p))
\end{equation*}
\noindent where $I(i) = d^i$ gives an exponentially decaying influence weight varying with fraction $d$ and walk length $i = 1 \cdots m$, and $W(p, \, L^i(p))$ denotes the number of walks from paper $p$ to each of the nodes reachable in $i$ steps (specifically, $L^i(p)$).  

Having established the \method for quantifying paper value according to impact, we can now pose the following (analogous) problems for authors and venues.

\begin{problem}[Author impact] 
{\bf Given:} the citation graph $G$, {\bf find:} a metric by which to quantify the value of authors according to their research impact.
\end{problem}

\begin{problem}[Venue impact] 
{\bf Given:} the citation graph $G$, {\bf find:} a metric by which to quantify the value of venues according to their research impact.
\end{problem}

We will treat these problems similarly using the derived \method for papers in order to demonstrate \emph{extensibility} to the broader author and venue contexts.  

We argue that the impact of an author is simply defined by the total impact of the works he produces.  Thus, we define the \method of an author $a$ as
\begin{align*}
\mathcal{S}(a) &= \sum_{p \in P^a} \mathcal{S}(p) \\
&= \sum_{p \in P^a} \sum_{i=1}^{m} \, I(i) \, W(p, \, L^i(p))
\end{align*}
\noindent where $P^a$ is the set of papers written by author $a$.  

In fact, venue impact can be defined in a similar way. Thus, we define the \method of a venue $v$ as
\begin{align*}
\mathcal{S}(v) &= \sum\limits_{p \in P^v} \mathcal{S}(p) \\
&= \sum_{p \in P^v} \sum\limits_{i=1}^{m} \, I(i) \, W(p, \, L^i(p))
\end{align*}
\noindent where $P^v$ is the set of papers published in venue $v$.  By this definition, the most impactful venues are those which have published the most impactful work.  Note that we do not take the mean over the number of papers as in JIF for 2 reasons: (a) the mean is a statistically inappropriate measure for heavy-tail distributions due to outlier sensitivity and (b) JIF reflects the \emph{reputation} of the journal rather than the impact (a venue which accepts very few articles with modest citation counts will often have a higher JIF than a journal which accepts more articles with a wider variety of citation counts).

We have now defined the \method for quantifying impact of papers, authors and venues.  \method can be \emph{interpreted} as a modified citation count which incorporates both direct and indirect impact.  It is worth noting that \method is meant to be construed as a ``lifetime achievement award,'' as it does not consider recency.  However, it offers very clear temporal extensions which can be useful in several scenarios.  We now present these.

In the paper ranking context, one can compute \method over the last $r$ years by using $L_r(H)$ instead of $L(H)$, where $L_r(H)$ only considers citations from works published within the previous $r$ years.  It follows naturally that all further descendants $L_r^k(H)$ for $k > 1$ were published in the last $r$ years as well, given the temporal connotation of citation edges.  This measure takes into account the recent impact of the paper in spurring new work in \emph{only} the last $r$ years -- it does not consider change in influence over the previous $r$ years from older citations, as incorporating this influence would unfairly bias the comparison towards older papers that had established many descendants.  We define the \methodr for papers as 
\begin{equation*}
\mathcal{S}_r(p) = \sum_{i=1}^{m} \, I(i) \, W(p, \, L_r^i(p))
\end{equation*}
For the author ranking context, we define $P_r^a$ instead of $P^a$, which contains only papers published by the author which have received citations in the previous $r$ years.  Again, it follows that if for all $p \in P_r^a$, $p$ was published in the last $r$ years, papers in $L_r^k(p)$ for $k \geq 0$ were also published in the last $r$ years.  This extension is especially useful, as one can easily compare how impactful two authors have been in recent years.  We define the \methodr for authors as
\begin{align*}
\mathcal{S}_r(a) &= \sum_{p \in P_r^a} \mathcal{S}(p) \nonumber \\
&= \sum_{p \in P_r^a} \sum_{i=1}^{m} \, I(i) \, W(p, \, L_r^i(p))
\end{align*}
In the same way, we can define $P_r^v$ instead of $P^v$ in the venue ranking context, which contains only papers published at that venue which received citations in the previous $r$ years.  The same principle for recency of the descendants of $p \in P_r^v$ holds as in the other contexts.  We define the \methodr for venues as
\begin{align*}
\mathcal{S}_r(v) &= \sum\limits_{p \in P_r^v} \mathcal{S}(p) \\
&= \sum\limits_{p \in P_r^v} \sum\limits_{i=1}^{m} \, I(i) \, W(p, \, L_r^i(p))
\end{align*}
Unlike the \method, the \methodr is not meant to be used for measuring overall impact.  Rather, it is an adaptation which accounts for \emph{temporality}.  In many cases, it is more relevant to examine recent performance and impact information rather than the aggregate, including recurring performance evaluation, ranking for modern relevance and comparison purposes.  Note that one could also additionally filter \methodr results to only include papers \emph{published within} rather than also \emph{receiving citations within} the recent $r$ years, to compare impact of only new papers if desired.  The former is a subset of the latter and can easily be computed post-hoc. 

\method and \methodr values can be quite large, especially for very influential papers, authors and venues.  For human parsing, we can scale the result into a comprehensible range by using $\overline{\mathcal{S}} = \log_2(\mathcal{S})$ and $\overline{\mathcal{S}_r} = \log_2(\mathcal{S}_r)$ in practice.  The logarithm function is monotonically increasing and will thus preserve ranking over the transformation. It further offers the attractive interpretable property of ``doubled impact'' for each additional point.

\subsection{Algorithm}
\label{sec:algo}

The last property which \method must satisfy is to be efficiently \emph{computable}.  It is clear that the most expensive component of the proposed computation is calculating the total number of walks of varying length from each paper.  While computing these values seems computationally daunting, it is not so in practice with careful design.  In fact, there exist much better solutions than the naive approach of counting walks on a per node basis via local graph search, which quickly becomes exponentially costly depending on path length and connectedness of $G$.  

One promising approach involves computing the number of walks in graph $G$ of varying length $i$ by taking powers of the adjacency matrix $\mathbf{A}$ of $G$.  It is well known that cell $(p_1,p_2)$ of $\mathbf{A}^i$ gives the total number of walks of length $i$ from $p_1$ to $p_2$.  We can next compute the row-sum for each $p_1$ to get the number of walks of length $i$ from $p_1$, and weight the result according to $I(i) = d^i$.  Moreover, we can iteratively compute $A^i$ by keeping only $A^{i-1}$ and $A$ in memory.  However, while sparse matrix multiplication is relatively efficient even for large matrices, memory constraints quickly become prohibitive given the increase in density of nonzeros for each additional exponentiation.  In our experiments on a machine with 400GB RAM, $\mathbf{A}^3$ cost roughly 110GB RAM to store.  The computation for $\mathbf{A}^4$ resulted in an out-of-memory error -- we expect the memory cost would over 1TB. 

\begin{algorithm}[t!] \DontPrintSemicolon
 \KwData{pap.-pap. adj. matrix $\mathbf{A}$, auth.-pap. adj. matrix $\mathbf{B}$, ven.-pap. adj.matrix $\mathbf{C}$, decay factor $d$, walk len. $m$}
 \KwResult{pap. scores $\mathbf{s}_p$, auth. scores $\mathbf{s}_a$, ven. scores $\mathbf{s}_v$}
 $\mathbf{s}_p = \mathbf{0}^T$ \tcp*[r]{dim $|P| \times 1$} 
 $\mathbf{s}_a = \mathbf{0}^T$ \tcp*[r]{dim  $|A| \times 1$} 
 $\mathbf{s}_v = \mathbf{0}^T$ \tcp*[r]{dim $|V| \times 1$}
 $\mathbf{v} = \mathbf{1}^T$ \tcp*[r]{dim $|P| \times 1$}
 \For{$i = 1$ \KwTo $m$}{
  $\mathbf{v} = \mathbf{A \cdot v}$\;
  $\mathbf{s}_p += d^i \cdot \mathbf{v}$\;
 }
 $\mathbf{s}_a = \mathbf{B} \cdot \mathbf{s}_p$\; 
 $\mathbf{s}_v = \mathbf{C} \cdot \mathbf{s}_p$\;
 \caption{\method}
 \label{alg:sindex}
\end{algorithm}

Fortunately, a more clever solution exists: instead of computing $\mathbf{A}^i$ and calculating the row-sum for each $p_1$, we can directly compute the total number of length $i$ walks by $\mathbf{A}^i \mathbf{v}$ where $\mathbf{v} = \mathbf{1}^T$ is the column 1-vector.  Thus, we avoid direct computation of $\mathbf{A}^i$ by instead computing $\mathbf{A(A(\ldots Av))}$ iteratively, in a manner similar to power iteration (though nonstochastic and unnormalized).  In each iteration, we compute $\mathbf{v} = \mathbf{Av}$ and thus maintain sparsity of $\mathbf{A}$.  We can then use a separate 0-vector $\mathbf{s}$ to accumulate \method scores in a single pass. The time complexity for each sparse matrix dense vector multiplication will be $O(|E_{pp}||P|)$, so with $m$ iterations we get $O(m|E_{pp}||P|)$ time-complexity which is linear on the number of papers (nodes), citations (edges) and walk length $m$, with only additional $O(|E_{pa}||A|)$ and $O(|E_{pv}||V|)$ for authors and venues respectively.  Furthermore, since we need to store only $\mathbf{A}$, $\mathbf{v}$ and $\mathbf{s}$ for each iteration, we can compute paper \method using a fixed space complexity of $O(|E_{pp}| \, + \,2|P|)$, with only additional $O(|E_{pa}| \, + \, |A|)$ and $O(|E_{pv}| \, + \, |V|)$ for author and venue \method respectively.  Algorithm \ref{alg:sindex} gives the concise algorithm.  To compute \methodr, we simply use the adjacency matrix $\mathbf{A}_r$ associated with the induced subgraph $G_r$, containing only edges from papers published in the last $r$ years -- complexity analysis is trivially similar. 
\section{Experiments}

In this section, we include qualitative and quantitative results from applying \method and \methodr on the Microsoft Academic Search (MAS) citation graph.  The graph consists of over \emph{119 million} papers, \emph{1 billion} citation edges, \emph{103 million} authors and \emph{21 thousand} venues -- for a more detailed description, we refer the reader to \cite{sinha2015overview}. We begin by first exploring some properties of \method in practice.  Next, we evaluate ranking correlation with traditionally used metrics and report \method and \methodr ranking results on the Microsoft Academic Search dataset.  Lastly, we discuss parameter selection and give results substantiating the scalability of our approach.

\subsection{Properties}


\subsubsection{Distribution}

\begin{figure*}[t!]
    \centering
    \begin{subfigure}[t]{0.32\linewidth}
        \centering
        \includegraphics[width=\linewidth]{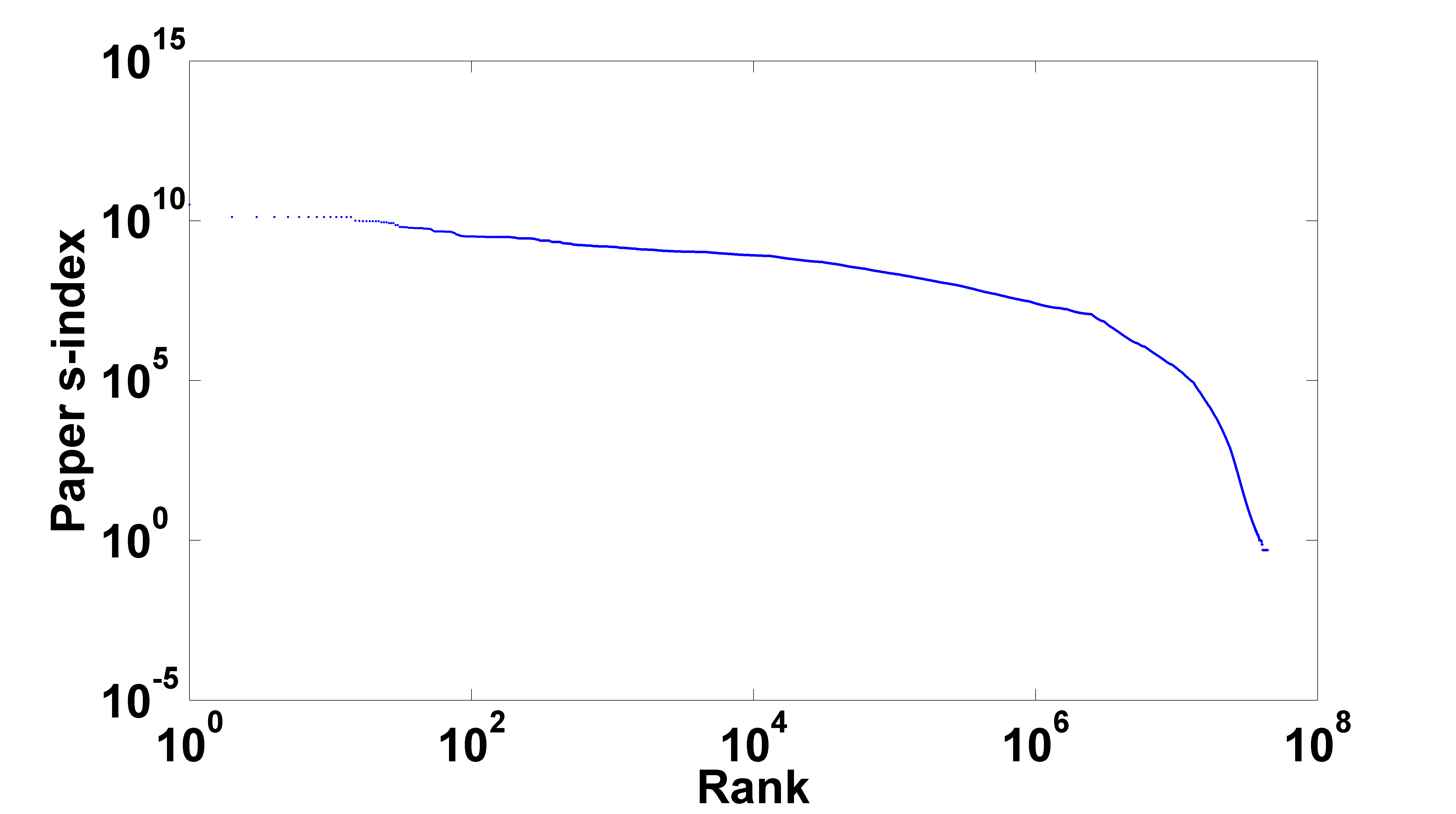}
        \caption{}
				\label{fig:pdist}
    \end{subfigure}%
    ~ 
    \begin{subfigure}[t]{0.32\linewidth}
        \centering
        \includegraphics[width=\linewidth]{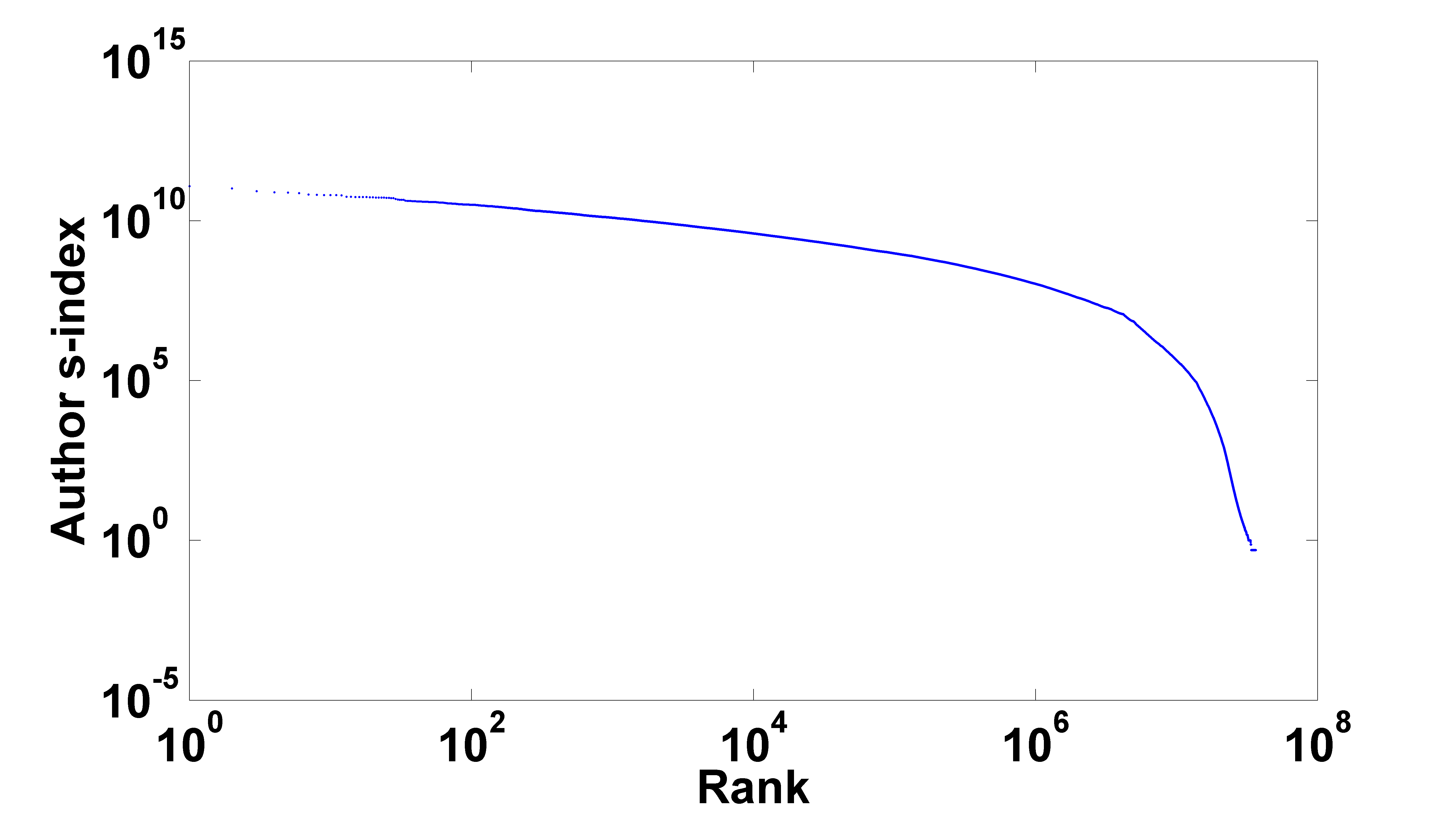}
        \caption{}
				\label{fig:adist}
    \end{subfigure}
		~
		\begin{subfigure}[t]{0.32\linewidth}
        \centering
        \includegraphics[width=\linewidth]{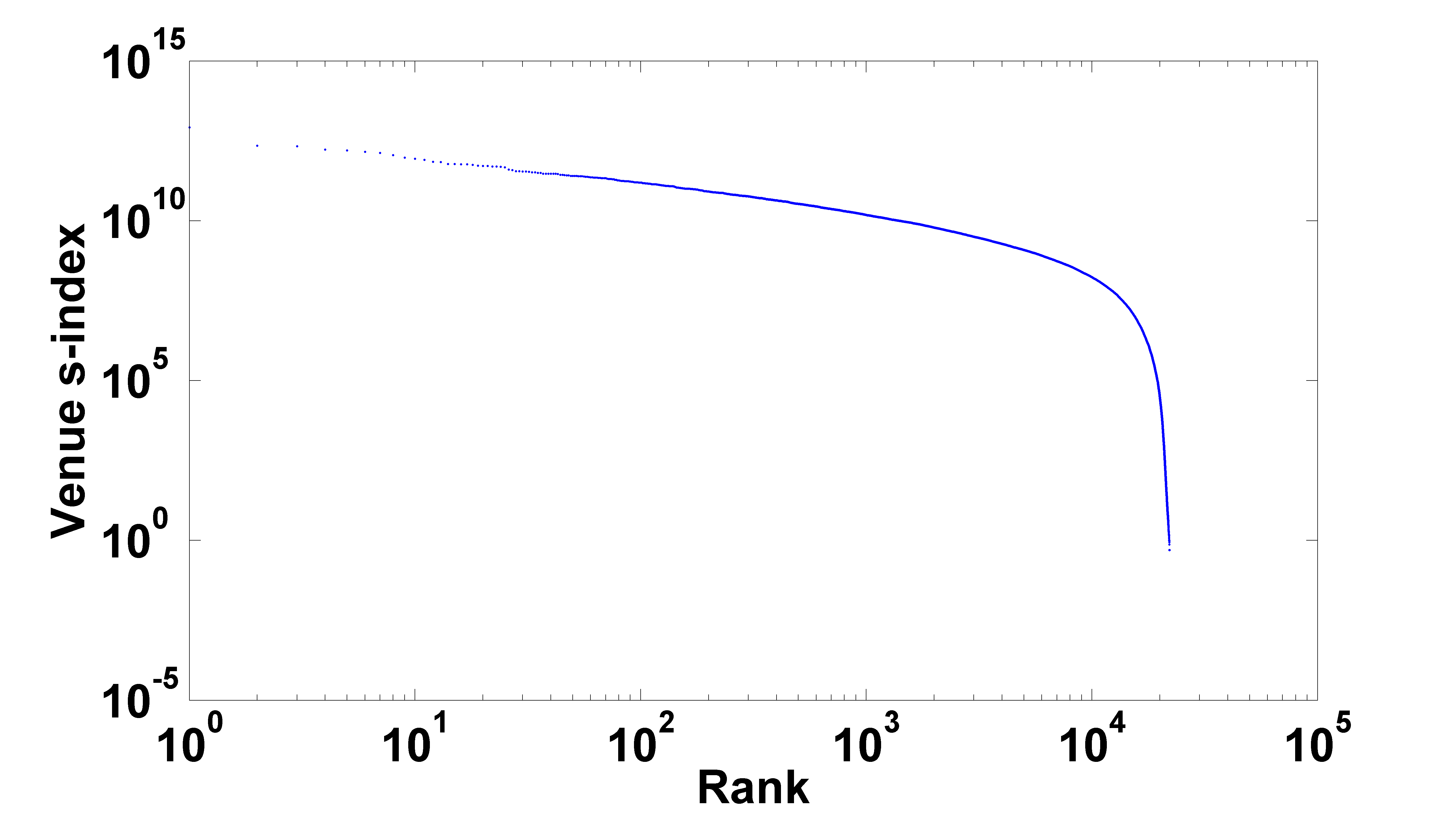}
        \caption{}
				\label{fig:vdist}
    \end{subfigure}
    \caption{Distribution of \method scores for (a) papers (b) authors and (c) venues.}
		\label{fig:sdistrib}
\end{figure*}

{\setlength{\tabcolsep}{7pt}
\begin{table*}[ht!]
\small
\centering
\caption{Top \method (top) and $s_5$-index (bottom) rankings on MAS data -- papers and authors selected across data mining/database/machine learning areas and venues across all areas.}
\label{tbl:masrankings}
\begin{tabular}{lll}
\toprule
\textbf{Papers} & \textbf{Authors} & \textbf{Venues} \\
\midrule \midrule
    Classification and Regression Trees                           & Robert E. Schapire  & Cancer                                          \\
    Basic Local Alignment Search Tool                                                      & Jiawei Han          & New England Journal of Med.             \\
    Occam's Razor                                                                          & Weiyin Loh          & Proc. of the Natl. Acad. of Sci. \\
    Pattern Recognition and Mach. Learning                                               & Michael Kearns      & The Lancet                                      \\
    The Strength of Weak Learnability                                                      & Stephen F. Altschul & Nature                                          \\
    C4.5: Programs for Mach. Learning                                                    & Webb C. Miller      & Science                                         \\
    The Nature of Statistical Learning Theory                                              & Warren Gish         & Arthritis and Rheumatism                        \\
    Neural Network Ensembles                                                               & David Lipman        & Journal of the Amer. Med. Assoc.     \\
    The Protein Data Bank                                                                  & Michael Stonebraker & Circulation                                     \\
    Advances in Knowledge Disc. and Data Mining                                        & Usama Fayyad        & Journal of Bio. Chem.                \\
    Genetic Alg. in Search Opt. and Mach. Learning                         & Christopher J. Merz & British Med. Journal                         \\
    The Comp. Complexity of Mach. Learning                                       & Rakesh Agrawal      & Cell                                            \\
    Efficient Distrib.-free Learning of Prob. Concepts                         & Manfred K. Warmuth  & Gastroenterology                                \\
    The Des. and Anal. of Efficient Learning Algorithms                               & Ramez Elmasri       & Blood                                           \\
    Separating Distrib.-free and Mist.-bound Learning Models & Christopher J. Date & Annals of Internal Med.                     \\
    Reliable Scheduling in a TMR Database System                                           & Sally A. Goldman    & Pediatrics                                      \\
    Learning Binary Relations and Total Orders                                             & Shamkant B. Navathe & Neurology                                       \\
    Mach. Learning: a Theoretical Approach                                               & Padhraic Smyth      & The Journal of Pediatrics                       \\
    On-line Learning of Linear Functions                                                   & Catherine Blake     & Annals of Neurology                             \\
    Learning Decision Trees Using the Fourier Spectrum                                     & John R. Quinlan     & The Amer. Journal of Med.                \\  
\midrule
      Basic Local Alignment Search Tool                      & Jiawei Han                & Cancer                                     \\
    Pattern Recognition in Mach. Learning                  & Usama Fayyad              & New England Journal of Med.                                    \\
    The Protein Data Bank                                  & Webb C. Miller            & Proc. of the Natl. Acad. of Sci.                 \\
    Classification and Regression Trees                    & Ramez Elmasri             &  The Lancet        \\
    Syst. and Int. Anal. of Gene Lists using DAVID    & Stephen F. Altschul       &  Nature                                 \\
    The Nature of Statistical Learning Theory              & Warren Gish               &   Science                                    \\
    Social Network Analysis: Methods and Applications      & Shamkant B. Navathe       &  Journal of the Amer. Med. Assoc.                                 \\
    Genetic Alg. in Search Opt. and Mach. Learning         & David J. Lipman           &   Circulation              \\
    Advances in Knowledge Disc. and Data Mining            & Eugene W. Myers           &    Journal of Bio. Chem                        \\
    Assoc. Rules and Data Mining in Hosp. Inf. Control     & Padhraic Smyth            &   Applied Physics Letters                     \\
    C4.5: Programs for Mach. Learning                      & Rakesh Agrawal            &   British Med. Journal                   \\
    Gene Expr. Omnibus: NCBI Gene Expr. Data Repo.    & Christopher J. Merz       &   Arthritis and Rheumatism                   \\
    The Elements of Statistical Learning                   & Gregory Piatetsky-Shapiro &    Cell                 \\
    Data Mining: Concepts and Techniques                   & Christopher M. Bishop     &    Journal of Applied Physics                                   \\
    Data Preparation for Data Mining                       & Philip S. Yu              &    Blood               \\
    NCBI GEO: Arch. for Genomic Data       & Nasser M. Nasrabadi       &   Annals of Internal Med.                                   \\
    Covering Numbers for Support Vector Mach.              & Christopher W. Clifton    &  Amer. Journal of Resp. Med.                   \\
    Maint. of Disc. Assoc. Rules in Large Databases        & Catherine Blake           &  The Amer. Journal of Med. \\
    Parallel Mining of Assoc. Rules                        & Christopher J. Date       &   Pediatrics                   \\
    CDD: A Cons. Domain Database for Inter. Anal. & John R. Quinlan           &  The Journal of Pediatrics                            \\
\bottomrule

\end{tabular}
\end{table*}

Figures \ref{fig:pdist}, \ref{fig:adist} and \ref{fig:vdist} show the distributions of \method scores across papers, authors and venues found in the MAS graph respectively.  The distributions are heavy-tailed, according to expectation, and suggest lognormal behavior -- few papers, authors and venues are extremely impactful, whereas the majority are less prolific.  Though the original citation count distribution is much closer to a power-law, the \method distribution becomes increasingly curved with greater walk-length $m$, as more and more low-cited papers are pushed to higher ranks due to indirect impact being accounted for.  

\subsubsection{Growth}

Although the growth over time of a paper, author or venue's \method depends entirely on how it impacts the scientific community, one might expect that the score for a popular paper would increase exponentially given the ``fan-out'' of the DAG rooted at a paper $p$, induced from $G$ by nodes in $\bigcup_{i = 1 \ldots m} L^i(\{p\})$ and the associated edges.
We find that for moderately and highly popular papers, exponential growth is indeed enjoyed for a time -- in fact, the full \method over time curves generally exhibit clear sigmoidal growth characterized by a period of dormancy, rapid direct and indirect citation and eventual taper.  The temporal length and rapidity of such growth are of course determined by innate popularity and contemporary relevance of the paper.  

Conversely, in cases where papers receive very few or no citations which are themselves poorly cited, the growth is better characterized as a step function in which changes to the \method happen sporadically over the years. This is characteristic of the famous ``diffusion of innovations'' theory proposed in \cite{rogers2010diffusion} which describes the process by which an innovation is communicated to participants in a social system over time.  The same sigmoidal diffusion pattern cannot be well observed for citation count, presumably because it only accounts for direct impact through citation.  

\subsection{Ranking Performance}

\subsubsection{Similarity to Existing Metrics}

\begin{figure*}[t!]
    \centering
    \begin{subfigure}[t]{0.32\linewidth}
        \centering
        \includegraphics[width=\linewidth]{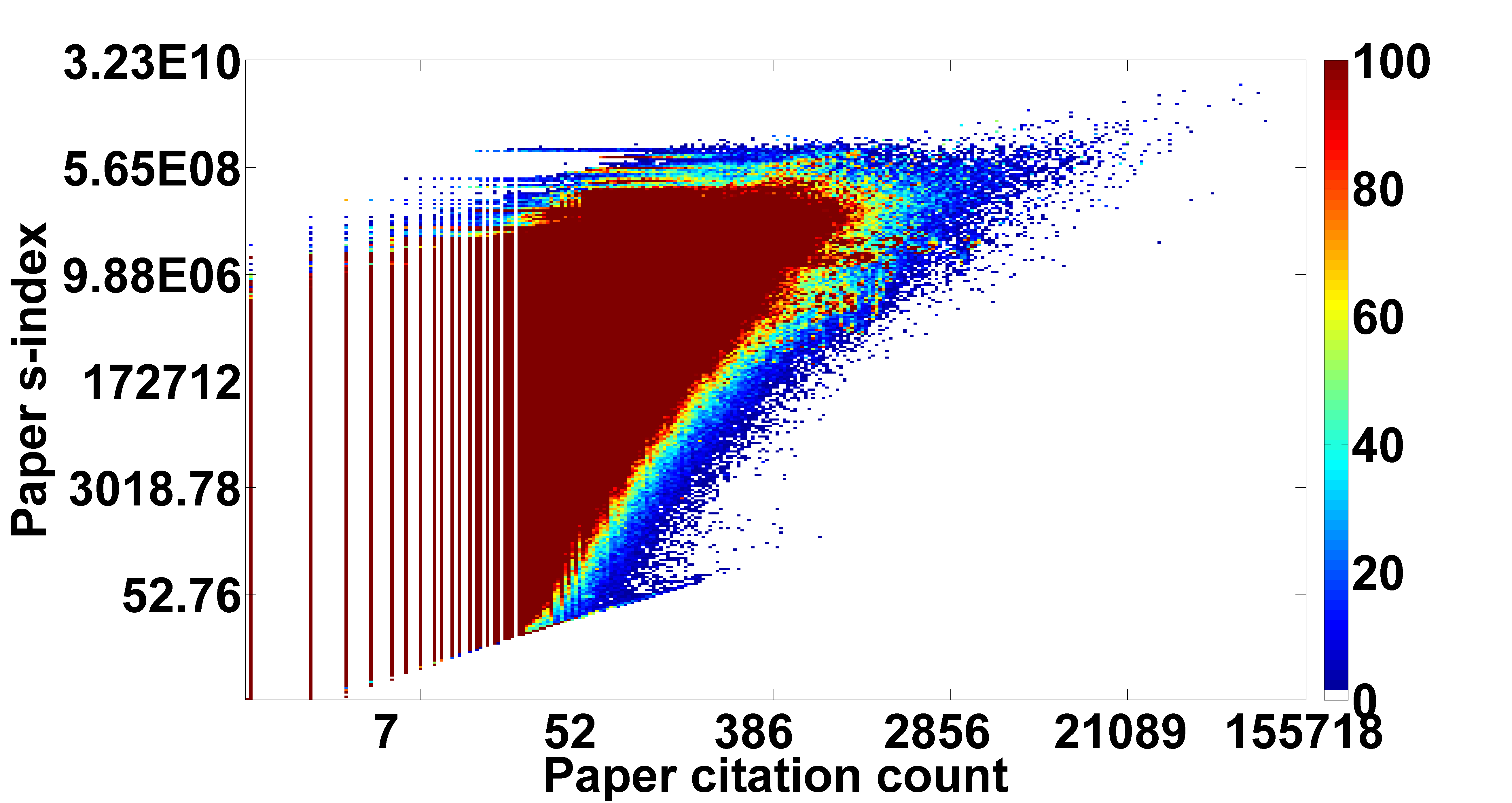}
        \caption{}
				\label{fig:vscitct}
    \end{subfigure}%
    ~ 
    \begin{subfigure}[t]{0.32\linewidth}
        \centering
        \includegraphics[width=\linewidth]{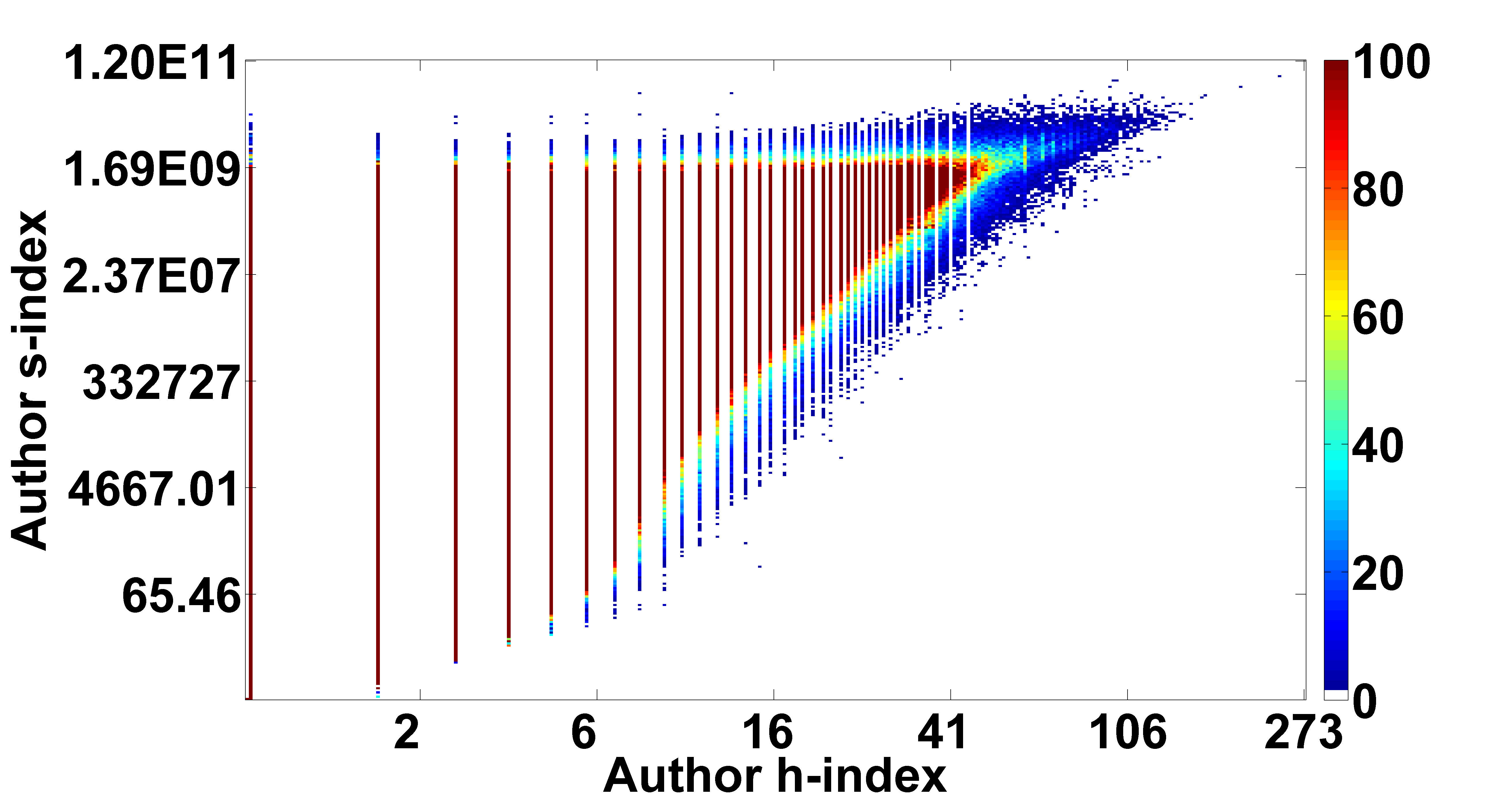}
        \caption{}
				\label{fig:vshindex}
    \end{subfigure}
		~
		\begin{subfigure}[t]{0.32\linewidth}
        \centering
        \includegraphics[width=\linewidth]{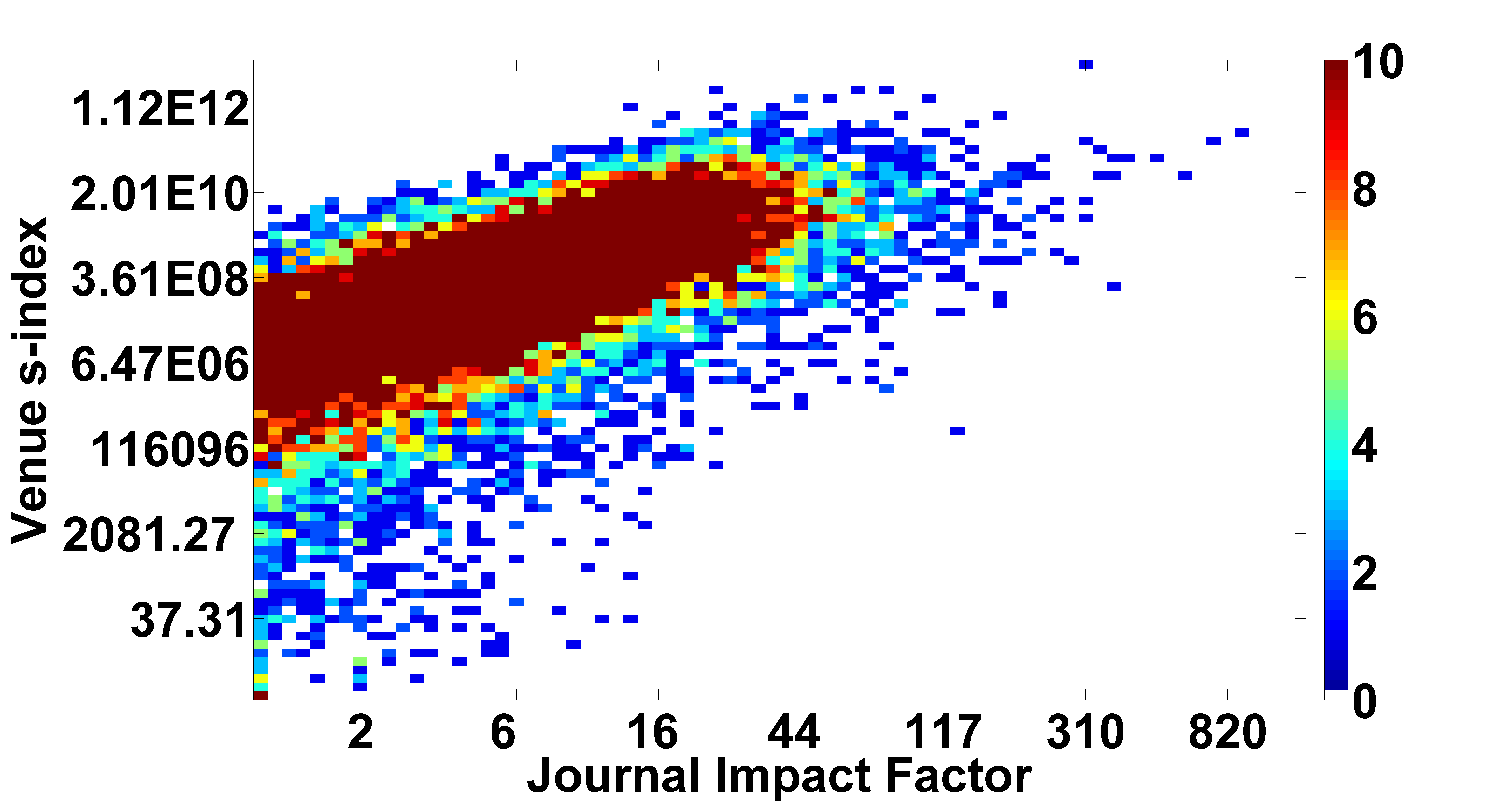}
        \caption{}
				\label{fig:vsjif}
    \end{subfigure}
    \caption{Correspondence between \method scores and (a) paper citation count, (b) author $h$-index and (c) venue JIF.  Colors denote density of points in logarithmically discretized bins in accordance with colorbars (red -- high, blue -- low).}
		\label{fig:compmetrics}
\end{figure*}

Figures \ref{fig:vscitct}, \ref{fig:vshindex} and \ref{fig:vsjif} show the relationship between \method and commonly used state-of-the-art metrics.  Correspondence with PageRank is not shown given the invalidity of the results for most papers given machine-precision issues (the overwhelming majority of papers have 0 impact and no meaningful ranking).

It is evident (and expected) that in all cases there are positive correlations between the respective \method and the metric scores.  Given that the Pearson correlation coefficient is ill-suited for tasks involving non-linear relationships, we use Spearman $\rho$ rank correlation coefficient defined as
\begin{equation*}
\rho = 1 - \frac{6 \sum d_i^2}{n(n^2 - 1)}
\end{equation*}
\noindent where $d_i = x_i - y_i$ is the difference between ranks and $n$ is the number of samples, in order to measure the strength of the relationships.  We find that $\rho = 0.78$ between \method and citation count, and $\rho = 0.49$ and $\rho = 0.76$ between \method and $h$-index and JIF respectively.  Perfectly correlated or inversely correlated ranking is characteristic of $\rho = 1$ and $\rho = -1$ respectively.  \cite{cohen2013statistical} notes that $\rho \geq 0.5$ is considered to be a ``large'' positive correlation.

Interestingly, despite the generally strong numerical correlations substantiated by the results in Figure \ref{fig:compmetrics}, it is apparent that there are many cases of poorly cited papers, low $h$-index authors and low JIF venues with \method scores characteristically higher than the norm and vice versa.  Further substantiating the value of measuring indirect impact via \method versus traditional ``direct'' metrics, we find that 57 of the 62 past Turing award winners can be found in the top $0.5\%$ of all authors ranked by \method as opposed to 50 when ranked by $h$-index -- a recall improvement of 11\%.

\subsubsection{Findings in Practice}

Table \ref{tbl:masrankings} shows the top 20 papers, authors and venues ranked using \method and \methodr ($r = 5$) on the MAS graph.  For paper and author ranking, we use an induced subgraph of papers and authors which have ``field of study'' labels corresponding to web mining, data mining, social network analysis, databases and machine learning.  For venue ranking, we use the entire graph containing all available data for papers.  We rank in this distinctive fashion to keep the discussion relevant to the reader, keeping in line with likelihood of expert familiarity.  We eliminate entries spuriously categorized into these fields as a result of the data collection process from the ranking for the same. Several of the top papers and authors in these rankings are well-known in the bioinformatics field, and appear because of association with the data mining ``field of study.''

Interestingly, several earlier foundational works ranked using \method disappear from the \methodr list, in favor of more up-and-coming and modern topics including social network analysis, applied data mining and bioinformatics.  Several of the authors also shift accordingly.  However, almost all of the venues remain the same, likely due to increased attraction due to tradition and established reputation.

\subsection{Parameter Selection}

\method is characteristic of two main parameters: the decay factor $d$ and the walk length $m$.  We select these parameters in a principled fashion, which we describe here.  

The decay factor $d$ is used to weight the influence of walks which are $i$ steps away.  It is similar to the damping factor used in PageRank, which describes the ``leakage probability'' of web surfers upon page visits.  Although PageRank uses a damping factor based on the observation that surfers typically follow on the order of 6 hyperlinks ($d = \frac{1}{6} \simeq 0.15$), \cite{chen2007finding} notes $d = 0.5$ is a better choice on citation graphs based on the frequency of feed-forward loops (see Figure \ref{fig:feedfwd}) in real data.  
Thus, we choose $d = 0.5$ to denote that 50\% of the influence of a paper on descendants is lost over each step.

$m$ denotes the maximum walk-length over which influence is computed.  We choose a small $m = 4$ in practice for multiple reasons: (a) the exponential influence decay will already heavily discount walks to ``far away'' papers -- $m = 4$ already produces a weight of only $\frac{1}{16}$, and (b) we expect that the content of far-away papers will lose relevance to the starting paper.  Moreover, we observe that the successive Spearman rank correlation $\rho$ rapidly approaches 1 after just a few steps -- between $m = 3$ and $m = 4$, $\rho$ is already $\geq 0.999$ with exponentially diminishing returns.  

\subsection{Scalability}

As described in Section \ref{sec:algo}, \method computation for papers is characterized by $O(m|E_{pp}||P|)$ time-complexity, which is linear on the number of papers (nodes), citations (edges) and walk length $m$.  Computing \method for authors and venues incurs small expenses of $O(|E_{pa}||A|)$ and $O(|E_{pv}||V|)$ operations respectively.  Figures \ref{fig:scal-m} and \ref{fig:scal-e} show linear scaling with respect to walk length $m$ and number of edges $|E_{pp}|$ for computing paper \method on the MAS graph using a MATLAB implementation.  We have additionally developed a Microsoft COSMOS (more generally, MS-SQL) implementation which runs on the MAS graph in minutes and is currently deployed and used regularly at Microsoft.   

\begin{figure}[t!]
    \centering
    \begin{subfigure}[t]{0.7\linewidth}
        \centering
        \includegraphics[width=\linewidth]{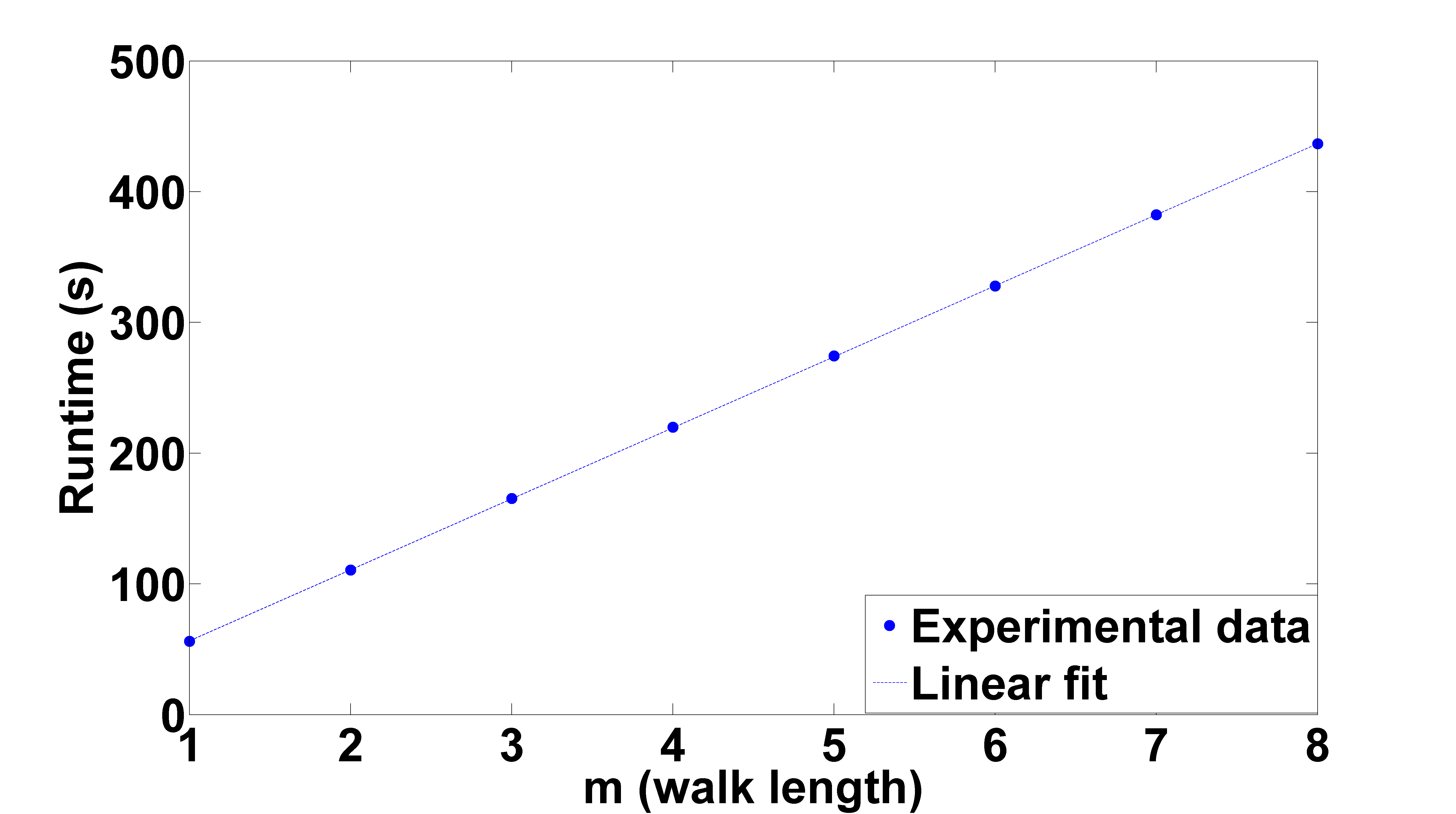}
        \caption{}
				\label{fig:scal-m}
    \end{subfigure}%
    \\ 
    \begin{subfigure}[t]{0.7\linewidth}
        \centering
        \includegraphics[width=\linewidth]{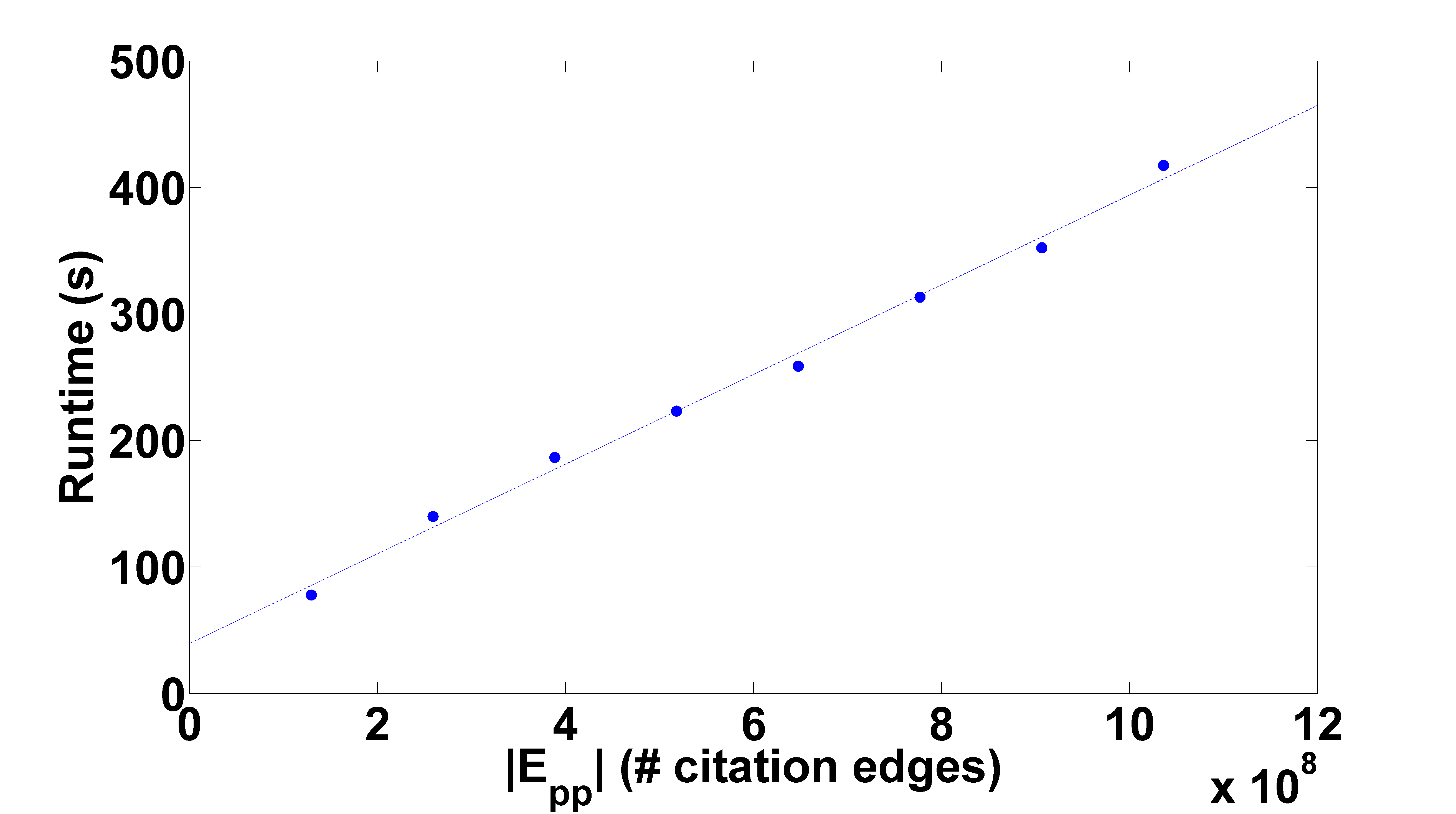}
        \caption{}
				\label{fig:scal-e}
    \end{subfigure}
    \caption{Linear scaling with respect to (a) walk length $m$ and (b) number of citation edges $|E_{pp}|$.}
\end{figure}
\section{Conclusion}

In this work, we aim to improve upon the state-of-the-art in impact metrics used for quantifying scientific research productivity.  While quantitative evaluation is by no means a functional replacement for carefully reading papers or qualitatively examining author contributions and evaluating peer-reviewed reputation, impact metrics are commonly used in managerial and strategic research decisions involving assigning tenure, awarding prizes, appointing academic posts, comparing researchers and deciding submission venues.  It is therefore important that these metrics be principled and behave according to human intuition.  In this work, we identify several desiderata that impact metrics should obey in practice and analyze how currently used state-of-the-art metrics violate these properties.  To this end, we next build towards the \method metric which quantifies impact of papers, authors and venues based on influence propagated over a citation graph and propose a fast, scalable algorithm for its computation which is currently deployed and used at Microsoft.  We evaluate \method on a large citation graph from Microsoft Academic Search and show promising results.

%
\bibliographystyle{abbrv}
\bibliography{bib/paper}  
%
%

\end{document}